\documentclass[prc,unsortedaddress,superscriptaddress,twocolumn,amssymb,showpacs,floatfix,nofootinbib]{revtex4}
\usepackage{graphics,epsfig}
\usepackage{amsmath}
\begin{document}

\date{\today}
\title{Three-body calculation of the rate of reaction $p+p+e \rightarrow d+\nu_e$ in the Sun }

\author{B.F. Irgaziev}\email{irgaziev@yahoo.com}
\affiliation{GIK Institute of Engineering Sciences and Technology,
Topi, Pakistan}
\author{V.B. Belyaev}
\affiliation{Joint Institute for Nuclear Research, Dubna, Russia}
\author{Jameel-Un Nabi}
\affiliation{GIK Institute of Engineering Sciences and Technology,
Topi, Pakistan}

\begin{abstract}
Using expansion of the three-body wave function of the $pep$
system in the initial state on hyperharmonic functions, the rate
of the $p+p+e^{-} \rightarrow d+\nu_e$ reaction in the Sun is
calculated. The results of calculation of the flux at 1 AU are
compared with the results of a measurement made by the Borexino
collaboration and Bahcall \textit{et al.} theoretical predictions.
\end{abstract}

\maketitle

\section{Introduction}
\label{int}

The study of the neutrino from the Sun is a multipurpose problem.
Indeed, the measurement of the solar neutrino flux incident on the
Earth helps to clarify the properties of the neutrino; for
example, the phenomenon of the oscillations and to determine its
parameters (the mixing angle and the eigenmass of the neutrino).
Then, the value of the flux and its spectral properties contain in
some cases information on the nuclear reactions  that in the
visible future cannot be observed in the laboratory. We mean here
the reactions:
\begin{equation}\label{pp}
p+p\rightarrow d+ e^{+} + \nu_e
\end{equation}
and
\begin{equation}\label{ppe}
p+p+e^{-}\rightarrow d+ \nu_e
\end{equation}
Moreover, as emphasized by John N. Bahcall in his book and in a
series of papers
\cite{bahcall,astro-ph/0209080,bahcall2004,bahcall-sab1,bahcall-sab2}
(see also \cite{www-sns-ias-edu}), solar neutrinos bring the
information about processes in the center of star connected with
solar structure models.

Although the reaction (\ref{ppe}), called the $pep$ reaction,
plays no significant role in hydrogen burning of stars, it is
essential in detecting  monoenergetic  neutrinos of
$E_{\nu}=1.442$ MeV. Also, a measurement of the neutrino flux from
reaction (\ref{ppe}) can be useful for the determination of the
parameters of the Standard Solar Model.

All the above considerations contain sets of parameters used for
fitting observable data in the framework of different models such
as star models, nuclear reactions models, and so on. An example of
such a type is the two-body model \cite{bahcall69} for reaction
(2) which is essentially three body \cite{belyaev}.

The purpose of this work is to exclude the model elements as much
as possible in the description of reaction (\ref{ppe}) and to
treat the initial state as a purely three-body state. From a
strict point of view, the modern treatment of any nuclear reaction
inside the Debye-sphere is still the model in a sense of the
absence of the dynamical consideration (it is the six-body problem
in the Sun's interior condition) of particles inside the sphere.
We hope, however, that the three-body instead of the two-body
treatment of reaction (\ref{ppe}) is a step in the right
direction.

Below, we concentrate ourselves mainly on the process (\ref{ppe})
for the following reasons.  First, in 2012 the results of the
first experimental observation of process (\ref{ppe}) was
announced \cite{borexino}, after more than 50 years of studies of
solar neutrino problems. The second reason is connected with the
absence of a three-body treatment of the initial state in process
(\ref{ppe}).

There is a question concerning sensitivity to the choice of the
nucleon-nucleon potential and related to these potentials wave
functions of the bound state of the deuteron and the continuum
state of the $pep$ system.

With all this in mind we present below the treatment which takes
into account all peculiarities of the $pep$ three-body system. In
Sec. \ref{input} we start from the inputs for the problem
considered: the weak Hamiltonian and $NN$-potentials, then in Sec.
\ref{gen}, we consider  the solution of the Schr\"odinger equation
to determine the $pep$ wave function of the initial state. In Sec.
\ref{sfac}, we present the calculation of the probability of the
$pep$ reaction and the astrophysical $S_{pep}$ factor taking into
account the Coulomb and strong interactions simultaneously. The
results for the rate of the process and fluxes of the neutrino are
discussed in Section \ref{rflux} and conclusions are presented in
Section \ref{fin}. The relevant Schr\"odinger equation for the
three interacting particles along with its hyperharmonics method
of solution is shown in Appendix A. We have used the program
{\small MATHEMATICA} (version 7) for our calculations.

\section{Inputs} \label{input}

The electron capture by the nuclear system can be described by the
following nonrelativistic effective weak Hamiltonian \cite{prim59}
\begin{eqnarray}\label{weakH}
H_w=\frac{1}{\sqrt{2}}\tau^{(+)}\frac{1-{\boldsymbol{\sigma\cdot\nu_1}}}{\sqrt{2}}\sum\limits_{i=1}^A
\tau_i^{(-)}\bigl[G_V\boldsymbol{1\cdot 1_i}+\nonumber\\
G_A\boldsymbol{\sigma\cdot\sigma_i}-G_P\boldsymbol{\sigma\cdot\nu_1\sigma_i\cdot\nu_1}\bigr]\delta{\boldsymbol{(r-r_i)}},
\end{eqnarray}
where $\boldsymbol{\nu_1}=\boldsymbol{\nu}/\nu$ ($\nu$ is the
neutrino momentum); $\boldsymbol{1}$,$\boldsymbol{1_i}$,
$\boldsymbol{\sigma}$ and $\boldsymbol{\sigma_i}$ are the $2\times
2$ matrix unit operators and spin angular momentum operators for
the lepton and \textit{i}th nucleon; $\boldsymbol{r}$ and
$\boldsymbol{r_i}$ are the space coordinates of the lepton and an
\textit{i}th nucleon; $\tau^{(+)},\,\,\tau^{(-)}_i$ are the
isobaric-spin operators which transfom a lepton electron state
into a lepton neutrino state and \textit{i}th nucleon proton state
into an \textit{i}th nucleon neutron state; and $G_V$, $G_A$ and
$G_P$ are the vector, axial vector and ``induced'' pseudoscalar
coupling constants, respectively. We take $G_V/(\hbar
c)^3=1.153\times 10^{-11}\,\rm{GeV}^{-2}$ and $G_A/(\hbar
c)^3=-1.454\times 10^{-11}\,\rm{GeV}^{-2}$ \cite{povh}. We can
simplify the weak Hamiltonian for the $pep$ system: the last term
in Eq. (\ref{weakH}) can be neglected because the emitted neutrino
has energy $E_\nu=1.442$ MeV and this term encloses factor
$\nu/2m_p$, where $m_p$ is the proton mass; the $pep\rightarrow
d+\nu_e$ transition satisfies the Gamow-Teller selection rule;
therefore, the first term of Eq. (\ref{weakH}) does not give
contribution to the matrix element of transition. Finally we take
into account that the electron neutrino has spin opposite to its
momentum $\nu$. Finally, for the weak Hamiltonian we get
\begin{equation}\label{Hweak}
H_w=\tau^{(+)}G_A\sum\limits_{i=1}^A\tau_i^{(-)}\boldsymbol{\sigma\cdot\sigma_i}\delta{\boldsymbol{(r-r_i)}}.
\end{equation}
With this Hamiltonian of Eq. (\ref{Hweak}) we obtain the electron
capture transition matrix element.

The energy of thermalization of particles in the interior of the
Sun corresponds to $E\sim 1.3$ keV which is small on the nuclear
energy scale. Therefore, we can use the simple $NN$-potential
which describes correctly the low energy data of the
nucleon-nucleon system. We apply the Gauss and the Yukawa
potentials \cite{blatt-jackson}. We fit the parameters of these
potentials to get the values of the deuteron energy, the
scattering lengths and the effective ranges for $pp$ and $pn$
scattering.

For the Gauss potential
\begin{equation}\label{gauss} V(r)=-V_{0} \exp \bigl(-r^{2}
/R_{N}^{2}\bigr).
\end{equation}
the calculation with the fitted parameters
$$ V_{0}^{s} ={\rm 30.36\; MeV,\; \; } R_{N}^{s} ={\rm 1.816\;
fm} $$ gives the scattering length $^sa_{pp}=-7.884$ fm and the
effective range $^sr_{pp}=2.678$ fm for $pp$ scattering  at the
singlet state $(s)$, while the parameters
$$ V_{0}^{t} ={\rm 60.572\; MeV,\; \; } R_{N}^{t} ={\rm 1.65\;
fm}, $$ lead to the scattering length $^ta_{np}=5.484$ fm,  the
effective range $^tr_{np}=1.85$ fm for $np$ scattering  at the
triplet state $(t)$ and the binding deuteron energy
$\varepsilon_d=2.225$ MeV.

The second potential is the Yukawa potential
\begin{equation}\label{yukawa} V(r)=-\frac{V_{0}}{r/R_N} \exp \bigl(-r
/R_{N}\bigr)
\end{equation}
with the parameters for the singlet state:
$$ V_{0}^{s} ={\rm 44.05\; MeV,\; \; } R_{N}^{s} ={\rm 1.206\;
fm}, $$ and for the triplet state:
$$ V_{0}^{t} ={\rm 53.27\; MeV,\; \; } R_{N}^{t} ={\rm 2.43\;
fm}. $$ The results of calculation of the scattering lengths and
effective ranges are:
\begin{eqnarray}^sa_{pp}&=&-7.782\,\,{\rm{fm}},\,\,\,\,
^sr_{pp}=2.868\,\,{\rm{fm}},\nonumber\\
^ta_{np}&=&\,\,\,\,\,5.626\,\,{\rm{fm}},\,\,\,\,
^tr_{np}=1.895\,\,{\rm{fm}}.
\end{eqnarray}

To find the neutrino flux we must use some Standard Solar Model
(SSM). There are several SSMs which are in good agreement with the
helioseismologically determined sound speed, temperature and
density of elements as a function of solar radius, the depth of
the convective zone, the surface helium abundance, and so on. We
applied data of parameters presented in the model BS05(OP)
\cite{bahcall95}. The results of Bahcall \textit{et al.}
\cite{bahcall-sab1} shows that the flux from the $pep$ reaction is
not sensitive to the type of SSM.

\section{The wave function of the $pep$ initial state}
\label{gen}

We note that Bahcall and May \cite{bahcall69} used the $pep$ wave
function in a factorized form as the product of the wave function
of the relative motion of two protons and the wave function of the
electron moving in the Coulomb field of these protons. However,
such a representation is not quite a correct procedure due to the
long-range nature of the Coulomb interaction, even for the
asymptotic behavior of the wave function when electron is at a
large distance from the protons \cite{alt93,muk96}. All the more,
we cannot perform such a factorization at a small relative
distance where we need to know the wave function with a sufficient
accuracy to get a good accuracy of the calculation for the
transition matrix element of the process $pep\to d\nu_e$. Also,
there is a problem with the total angular momentum of $pep$,
because the moments of subsystems are not conserved. Fortunately,
the relative angular moment of two nucleons in the initial and
final states is zero as angular moments of the electron and the
neutrino; therefore, the last problem did not arise in the Bahcall
calculations of the $pep$ reaction.

We use the hyperspherical harmonics \cite{djibuti,fabre} for
expansion of the $pep$ wave function in the initial state and
solve directly the three-body Schr\"odinger equation and,
therefore, we are free of the problems which we mentioned above.
As in the nonrelativistic approach the orbital and spin moments
are conserved independently of one another, we can expand the
spatial part of the $pep$ three-particle wave function over
hyperspherical functions and we obtain the linked system of the
radial differential equations. Derivation of the system radial
equations is given in  Appendix \ref{app1}. To proceed further, we
now use the system of Eqs. (\ref{rad eq}) for the radial wave
functions. Since the total energy of the $pep$ system  is low, the
main contribution to the three-particle wave function gives the
states with the zero relative orbital moments. Owing to the
centrifugal potential and $\kappa\rho\ll 1$ at small distance
where we need to calculate the wave function with higher accuracy,
contributions of the components with the hypermoments $K>0$ should
be suppressed in the total wave function. Taking into account
these conditions, we need to find the solution of the single
equation only for $K=l_x=l_y=0$ and with  the nondiagonal terms
omitted. We omit all the indices because they correspond to zero
values of the quantum numbers and get the equation for the radial
wave function $U(\rho)$:
\begin{equation}\label{rad eq0}
\frac{d^2U(\rho)}{d\rho^2}+\frac{1}{\rho}\frac{dU}{d\rho}-\Bigl({\cal{V}}(\rho)+\frac{4}{\rho^2}-\kappa^2\Bigr)U(\rho)=0,
\end{equation}
where $\kappa^2=2\mu_{23}E/\hbar^2>0$ ($E$ is the total energy of
the $pep$ system);
\begin{eqnarray}\label{Vpot0}
{\cal{V}}(\rho)\,\,\,\,&=&{\cal{V}}^N(\rho)+{\cal{V}}^C(\rho),\\
{\cal{V}}^C(\rho)&=&\frac{32\mu_{23}}{3\pi\hbar^2}\frac{(a_1+a_2+a_3)}{\rho}\equiv\frac{2\eta_3\kappa}{\rho}.
\end{eqnarray}
Here $\eta_3$ is the three-body Coulomb parameter which is defined
as
\begin{eqnarray}\label{eta}
\eta_3&=&\frac{16\mu_{23}}{3\pi\hbar^2\kappa}(a_1+a_2+a_3),\\
a_1&=&\sqrt{\frac{m_2m_3}{\mu_{23}(m_2+m_3)}}e^2\equiv e^2,\label{a1}\\
a_2&=&-\sqrt{\frac{m_1m_3}{\mu_{23}(m_1+m_3)}}e^2\simeq
-\sqrt{\frac{2m_1}{m_3}}e^2,\label{a2}\\
a_3&=&-\sqrt{\frac{m_1m_2}{\mu_{23}(m_1+m_2)}}e^2\simeq
-\sqrt{\frac{2m_1}{m_2}}e^2.\label{a3}
\end{eqnarray}
The matrix element of the nuclear potential ${\cal{V}}^N$ is the
following: for the Gauss potential
\begin{equation}
{\cal{V}}^N(\rho)=-\frac{8\mu_{23}V_0}{\hbar^2}\frac{\exp\Bigl(-\frac{\rho^2}{2R_N^2}\Bigr)I_1\Bigl(\frac{\rho^2}{2R_N^2}\Bigr)}
{\rho^2/R_N^2},
\end{equation}
and for the Yukawa potential
\begin{equation}
{\cal{V}}^N(\rho)=-\frac{16\mu_{23}V_0}{\hbar^2}\,\frac{\frac{2\rho}{3\pi
R_N}- I_2(\frac{\rho}{R_N})+L_2(\frac{\rho}{R_N})}{\rho^2/R_N^2},
\end{equation}
where $I_n(z)$ is the modified Bessel function of the first kind,
and $L_n(z)$ is the modified Struve function.

To find a unique solution of Eq. (\ref{rad eq0}) for the
continuous state of the $pep$ system, we must determine boundary
conditions. Instead of defining the function $U(\rho)$ and its
derivative at the origin ($\rho=0$), we define the wave function
at a point $\rho_0$ close to zero because we know the behavior of
the wave function near the origin. At small distances from the
origin they have a form \footnote{We take the point $\rho_0$ not
equal to zero for reasons of the numerical calculations.}.
\begin{equation}\label{b.con}
U(\rho_0)=J_2(\kappa_0 \rho_0),\,\,\,\, U^\prime (\rho_0)=\kappa_0
J_2^\prime(\kappa_0 \rho_0),
\end{equation}
where $\kappa_0=\sqrt{\kappa^2+\mid{\cal{V}}_N(\rho_0)\mid}$.

The wave function $U(\rho)$ at large distance, where the nuclear
interaction is negligible, has the following asymptotics:
\begin{equation}\label{assCoul}
U(\rho)\xrightarrow{\rho \to
\infty}e^{i\delta_3}\cos\delta_3\bigl(F_{00}(\kappa\rho)-\tan\delta_3G_{00}(\kappa\rho)\bigr),
\end{equation}
where $\delta_3$ is the three-body nuclear scattering  phase shift
modified by the Coulomb interactions, $F_{00}(\kappa\rho)$ and
$G_{00}(\kappa\rho)$ are the three-body regular  and irregular
Coulomb wave functions, respectively, and are defined as
\begin{eqnarray}
F_{00}(\kappa\rho)&=&\frac{1}{2}\sqrt{\frac{2}{\pi \kappa\rho}}e^{\frac{\pi }{2}\eta_3}\left[e^{i(\delta_{3C}-\frac{5}{4}\pi)}W_{-i\eta_3, 2}(-2i \kappa\rho)+\right.\nonumber\\
&&\left. e^{-i(\delta_{3C}-\frac{5}{4}\pi)}W_{i\eta_3, 2}(2i
\kappa\rho)\right],\\
G_{00}(\kappa\rho)&=&\frac{1}{2}\sqrt{\frac{2}{\pi \kappa\rho}}e^{\frac{\pi }{2}\eta_3}\left[e^{i(\delta_{3C}-\frac{5}{4}\pi)}W_{-i\eta_3, 2}(-2i \kappa\rho)-\right.\nonumber\\
&&\left. e^{-i(\delta_{3C}-\frac{5}{4}\pi)}W_{i\eta_3, 2}(2i
\kappa\rho)\right],
\end{eqnarray}
where $\delta_{3C}$ is the three-body Coulomb phase shift given by
\begin{equation}
\delta_{3C}=\arg\left[\Gamma(5/2+i\eta_3)\right],
\end{equation}
and $W_{\lambda, \mu}(z)$ is the Whittaker  function.  In the
numerical calculations, when we are dealing with large values of
the Coulomb parameter $\eta_3$ and $\kappa\rho<1$, it is best to
use another representation of the function $F_{00}$:
\begin{eqnarray}
F_{00}(\kappa\rho)&=&\frac{1}{2}\sqrt{\frac{2}{\pi\kappa\rho}}\,
\frac{\mid\Gamma\left(\frac{5}{2}+i\eta_3\right)\mid}{4!}e^{-\frac{\pi\eta_3}{2}-i\kappa\rho}\times\nonumber\\
&&\,{_1}F_1\left(\frac{5}{2}-i\eta_3;5;i2\kappa\rho\right),
\end{eqnarray}
where $\,_1F_1(a;b;z)$ is the confluent hypergeometric function.

We solve Eq. (\ref{rad eq0}) using the boundary conditions
(\ref{b.con}) and then matching the logarithmic derivative of the
solution in the asymptotic region with the logarithmic derivative
of the asymptotic solution [Eq. (\ref{assCoul})] we define the
three-body phase shift $\delta_3$ which depends on the total
energy $E$ of the $pep$ system. The numerical calculations with
the Gauss potential yield the following values of the phase
shifts: at the energy $E = 1.4$ keV corresponding to the
temperature at the core of the Sun, we obtain $\delta_3=3.7\times
10^{-16}$ radian, while at  $E = 6.0$ keV it is $2.3\times
10^{-9}$ radian. Even with energy as high as 20 keV, the values of
the phase shift remains very small. We get the same results with
the Yukawa potential.

Matching the numerically obtained solution of the Schr\"odinger
equation with its asymptotics allows us also to find the
normalization factor. At large distances the ratio of the
unnormalized solution to the asymptotic function becomes a
constant.

\begin{figure}
\epsfig{file=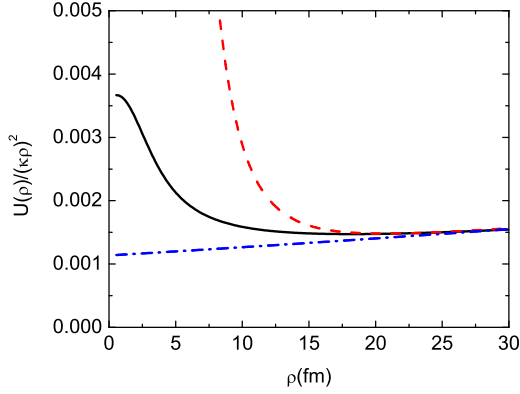,width=7.8cm} \caption{(Color
online) The $pep$  radial wave function  (solid curve) obtained by
solving Eq. (\ref{rad eq0}) and the function defined by
Eq.(\ref{assCoul}) (dashed curve). The dot-dashed curve shows the
pure Coulomb function  $F_{00}(\kappa\rho)$. All presented
functions are divided by $(\kappa\rho)^2$.
$E_{pep}=6\,\,\rm{keV}$. \label{fig1}}
\end{figure}

Figure \ref{fig1} shows the results of calculation of the $pep$
radial wave function (solid curve) at $E=6$ keV and its
asymptotics (\ref{assCoul}) for the Gauss $NN$ potential (dashed
curve). Since the scattering phase shift $\delta_3$ is close to
zero at the considered energy range the pure three-body Coulomb
wave $F_{00}(\kappa\rho)$ (dot-dashed curve) is close to the
solution of the Schr\"odinger equation at the distances $\rho
> 30$ fm. Note that all functions presented on Figure \ref{fig1} are
divided by $(\kappa\rho)^2$. As seen in Figure \ref{fig1}, the
function $F_{00}(\kappa\rho)/(\kappa\rho)^2$ has a linear
dependence on $\rho$ at the considered range of the variable
$\rho$. We obtain the same results for the Yukawa $NN$ potential.
We need to know the three-body wave function with high accuracy in
the interval $0<\rho<35$ fm because the deuteron wave function
cuts integration at the distance 35 fm when we calculate the
matrix element of the $pep\to d+\nu$ transition.

\section{The probability of the $pep\to d+\nu$ reaction and astrophysical $S$ factor}
\label{sfac}
 Using the weak Hamiltonian (Eq. (\ref{Hweak})) we
can write the matrix element of the $pep$ reaction as
\begin{eqnarray}\label{matrix}
H_{if}&=& G_A
<\varphi_\nu\mid{\boldsymbol{\sigma}}\tau^{(+)}\mid\varphi_e>\times\nonumber\\
&&\sum\limits_{i=1}^{2}<\Psi_d\Psi_\nu\mid{\boldsymbol{\sigma}}_i\tau_i^{(-)}\mid\Psi_{pep}>,
\end{eqnarray}
where $\varphi_e$ and $\varphi_\nu$ are the spin functions of the
electron and neutrino, respectively; $\Psi_d$ is the wave function
of the deuteron, while $\Psi_{pep}$ is the three-body wave
function of the $pep$ system in a continuous state. The neutrino
wave function $\Psi_\nu$ can be taken as a plane wave, because it
does not interact with the deuteron. At the energy of the emitted
neutrino is 1.44 MeV and the de Broglie wavelength of the neutrino
is almost 900 fm, implying the plane wave is essentially unity
over the effective volume of the deuteron (the difference from
unity is about 0.2\%). Accordingly, we take $\Psi_d=1$. In the
wave function $\Psi_{pep}$ the electron coordinate is taken at the
point where one of the protons is placed.

Since the weak interaction is small, first-order perturbation
theory can be applied to calculate of the probability $P_3$ of
reaction per unit of time; therefore, we have
\begin{equation}\label{prob}
P_3=\frac{2\pi}{\hbar}\overline{\mid H_w\mid^2}\rho(E_\nu),
\end{equation}
where the overline means summation over spins in  the final state
and averaging over spins in the initial state, and the density of
neutrino states is
\begin{equation}
\rho(E_\nu)=\frac{E_\nu^2}{2\pi^2\hbar^3c^3}.
\end{equation}
Here $E_\nu$, and $c$ are the neutrino energy and light speed,
respectively. Using the energy conservation law we define
\begin{equation*}
E_\nu=E_{pp}+E_e-(M_d -2m_p-m_e)c^2,
\end{equation*}
where $E_{pp}$ is the kinetic energy of proton-proton relative
motion, $M_d,\,\,m_p,\,\,m_e$ are the deuteron, proton and
electron masses, respectively. If we put kinetic energies of the
particles to zero, we get $E_\nu=1.442$ MeV.

Averaging in  Eq. (\ref{prob}) and using the Jacobi coordinates
for the particles defined by Eq. (\ref{jacob}) of Appendix
\ref{app1} we obtain
\begin{equation}
P_3=\frac{3E_\nu^2G_A^2}{\pi\hbar^4c^3}\mid\int\Psi^*_d({\boldsymbol{x_1}})
\Psi_{pep}({\boldsymbol{x_1,y_{10}}})d^3x_1\mid^2.
\end{equation}
Note that the presence of the delta-function in the weak
Hamiltonian (\ref{Hweak}) leads to calculation of the integral
over the variable $\boldsymbol{y_1}$ to the value of the $pep$
wave function at the distance
${\boldsymbol{y_{10}}}=\sqrt{\frac{m_1(m_2+m_3)}{\mu_{23}(m_1+m_2+m_3)}}{\boldsymbol{x_1}}/2\simeq
\sqrt{\frac{m_e}{2m_p}}{\boldsymbol{x_1}}$.

The deuteron wave function can be written in the form
\begin{equation}\label{deut-wave}
\Psi_d({\boldsymbol{x_1}})=\frac{\chi_d(x_1)}{x_1}Y_{00}({\boldsymbol{\hat{x}_1}})\equiv
\frac{1}{\sqrt{4 \pi}}\frac{\chi_d(x_1)}{x_1},
\end{equation}
where the radial wave function $\chi_d(x_1)$ is normalized so that
the deutron wave function $\Psi_d({\boldsymbol{x_1}})$ is
normalized to unity.

Using the first term of  Eq. (\ref{expansion}) from Appendix
\ref{app1} and taking into account that $K=l_x=l_y=0$ the
three-body wave function of the $pep$ system in the initial state
can be written as
\begin{equation}\label{Psi0}
\Psi_{pep}({\boldsymbol{x_1},\boldsymbol{y_1}})= \frac{8\,
U(\rho)}{(\kappa\rho)^2}
\end{equation}
Taking into account the last two equations, we obtain that the
overlap integral is
\begin{eqnarray}\label{overlap}
\int\Psi_d^*({\boldsymbol{x_1}})\Psi_{pep}({\boldsymbol{x_1,y_{10}}})d^3x_1=\,\,\nonumber\\
\frac{16\sqrt{\pi}}{\kappa^2}\int\limits_0^\infty
x_1^{-1}\chi_d(x_1)U(x_1)dx_1.
\end{eqnarray}
Here the three-body radial wave function is calculated at the
point $\rho=\sqrt{x_1^2+y_{10}^2}\simeq x_1$.

\begin{figure}
\epsfig{file=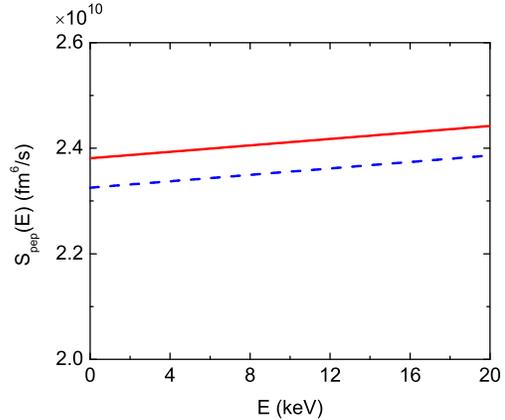,width=7.8cm} \caption{(Color online)
The $pep$ astrophysical $S_{pep}$ factor  as a function of energy
calculated by the Gauss (solid line) and Yukawa (dashed line) $NN$
potentials.
 \label{fig2}}
\end{figure}

Usually in the nuclear astrophysics, for parameterizations of a
two-body reaction cross section the astrophysical $S$ factor is
used \cite{angulo}. To find the $S$ factor from the cross section,
the Gamow factor is extracted from the cross section, i.e.
\begin{equation}\label{S_pp}
\sigma(E)=\frac{S(E)}{E}e^{-2\pi\eta},
\end{equation}
where $\eta=Z_1Z_2e^2/(\hbar v)$ is the Sommerfeld (Coulomb)
factor for the colliding nuclei with charges $Z_1$ and $Z_2$
having the relative velocity $v$. A similar procedure can be
carried out for the reaction with three particles in the initial
state because the three-body radial wave function $U(\rho)$ of the
continuum state at a distance larger than the nuclear interaction
radius encloses factor $\exp(-\pi\eta_3/2)\mid\Gamma(5/2 +
i\eta_3)\mid$. Therefore we define the astrophysical $S_{pep}$
factor for the $pep$ reaction as
\begin{eqnarray}\label{S_3}
P_3(E)&=&G_0(E)S_{pep}(E),\\
G_0(E)&=&\frac{2\pi
e^{-2\pi\eta_3}(\frac{1}{4}+\eta_3^2)(\frac{9}{4}+\eta_3^2)}{1+e^{-2\pi\eta_3}}.\label{gamow_3}
\end{eqnarray}
$S_{pep}(E)$ is almost  a linearly varying function of $E$ and
$S_{pep}(0)$ is not equal to zero, like the astrophysical $S_{pp}$
factor for two-body reactions. Note that the unit of $S_{pep}$
coincides with the unit of $P_3$ because $G_0(E)$ is dimensionless
and it is the Gamow factor for the $pep$ reaction. If we use
expansion of $S_{pep}(E)$ over $E$ as
\begin{equation}\label{expand}
S_{pep}(E)=S_0+S_1 E+S_2E^2,
\end{equation}
we obtain the following results for the value of the
coefficients:\\
For the Gauss potentials
\begin{eqnarray}\label{coef-gauss}
S_0&=&2.38\times 10^{10} \rm{fm^6/s},\nonumber\\
 S_1&=&3.03\times
10^{10}
\rm{fm^6/(MeV\,s)}, \nonumber\\
S_2&=&1.45\times 10^{10} \rm{fm^6/(MeV^2\, s)},
\end{eqnarray}
and for the Yukawa potentials
\begin{eqnarray}\label{coef-yukawa}
S_0&=&2.33\times 10^{10}\, \rm{fm^6/s};\nonumber\\
 S_1&=&3.01\times
10^{10}\,
\rm{fm^6/(MeV\, s)};\nonumber\\
S_2&=&1.78\times 10^{10}\,\rm{fm^6/(MeV^2\, s)}.
\end{eqnarray}
Behavior of the $S_{pep}$ astrophysical factor on energy is
presented in Fig. \ref{fig2} where linear dependence of the $S$
factor is seen clearly for both potentials. The small difference
between the $S_{pep}$ factors for the Gauss and Yukawa potentials
can be explained by small differences of the obtained low energy
scattering parameters.
To check the validity of the results of our calculations of the
$pep$ reaction, we calculated  the astrophysical $S_{pp}$ factor
for the $pp$ reaction, according to the conventional  definition
(Eq. (\ref{S_pp})) by the same $NN$ potentials. If $S_{pp}$ is
approximated by a polynomial
\begin{equation}\label{SS-pp}
S_{pp}(E)=S_0+S_1E+S_2E^2,
\end{equation}
we get the following results:\\
For the Gauss potentials
\begin{eqnarray}\label{pp-gauss}
S_0&=&4.21*\times 10^{-25} \rm{MeV\, b};\nonumber\\
 S_1&=&4.75\times
10^{-24}
\rm{b}; \nonumber\\
S_2&=&3.02\times 10^{-23} \rm{MeV^{-1}\, b};
\end{eqnarray}
and for the Yukawa potentials
\begin{eqnarray}\label{pp-yukawa}
S_0&=&4.11\times 10^{-25}\, \rm{MeV\, b};\nonumber\\
 S_1&=&4.64\times
10^{-24}\,
\rm{b};\nonumber\\
S_2&=&2.93\times 10^{-23}\,\rm{MeV^{-1}\, b}.
\end{eqnarray}
These results are very close to data presented in \cite{angulo}.

\section{Rate of the $pep$ reaction and the solar neutrino flux}
\label{rflux}
We introduce  the \textit{rate constant} of the
$pep$ reaction as
\begin{eqnarray}\label{rate}
{\cal{K}}_{pep}=<P_3>=\,\,\,\,\,\,\,\,\,\,\,\,\,\,\,\,\,\,\,\,\,\,\,\,\,\,\,\,\,\,\,\,\,\,\,\,\,\,\,\,\,\,\,\,\,\,
\,\,\,\,\,\,\,\,\,\,\,\,\,\,\,\,\,\nonumber\\
\,\,\,\,\int\limits_0^\infty\int\limits_0^\infty\int\limits_0^\infty\varphi(v_e)\varphi(v_{p_1})\varphi(v_{p_2})P_3(E)dv_edv_{p_1}dv_{p_2},
\end{eqnarray}
where $\varphi(v)=4\pi v^2(\frac{m}{2\pi
kT})^{3/2}\exp\Bigl(-\frac{mv^2}{2kT}\Bigr)$ is the
Maxwell-Boltzmann distribution function for particles of mass $m$
and velocity $v$ and $k$ is the Boltzmann constant. Excluding the
velocity of the center of mass of the $pep$ system and taking into
account the fact that the center of mass of the $pep$ system is
almost the same as the center of the $pp$ system we get
\begin{equation}\label{rate-ppe}
{\cal{K}}_{pep}=\int\limits_0^\infty\int\limits_0^\infty\varphi(v_e)\varphi(v_{pp})P_3(E)dv_edv_{pp},
\end{equation}
where $\varphi(v_{pp})=4\pi v_{pp}^2(\frac{\mu_{pp}}{2\pi
kT})^{3/2}\exp\Bigl(-\frac{\mu v^2}{2kT}\Bigr)$, $v_{pp}$ is the
relative velocity of protons and $\mu_{pp}=m_p/2$. Taking into
account that
$$E=\frac{\mu_{pp}v_{pp}^2}{2}+\frac{m_ev_{e}^2}{2}$$
and making transformations
$$v_{pp}=V\cos\alpha,\,\,\,\,\,
v_e=\sqrt{\frac{\mu_{pp}}{m_e}}V\sin\alpha,\,\,\,\,E=\frac{\mu_{pp}V^2}{2}$$
we obtain
\begin{equation}\label{rate-fin}
{\cal{K}}_{pep}=\frac{1}{(kT)^3}\int\limits_0^\infty
e^{-E/kT}P_3(E)E^2dE.
\end{equation}
Using Eqs. (\ref{S_3}), (\ref{gamow_3}) we finally obtain
\begin{equation}\label{r-k}
{\cal{K}}_{pep}=\frac{1}{(kT)^{3}}\int\limits_0^\infty
G_0(E) S_{pep}(E) e^{-E/kT}E^2dE.
\end{equation}
We note that \textit{the rate constant} depends on the temperature
and the nature of the reactants, but does not depend on their
concentration. If we define the Gamow energy as
\begin{equation}\label{G-w-e}
E_G=\frac{1}{2}\mu_{23}\Bigl(\frac{2\pi e^2
z_{eff}}{\hbar}\Bigr)^2,
\end{equation}
the integrand of Eq. (\ref{r-k}) is a maximum at the energy
\begin{equation}\label{max-E}
E_{max}=\Bigl(\frac{1}{2}kT\sqrt{E_G}\Bigr)^{2/3}.
\end{equation}
Here $z_{eff}$ is defined through Eqs. (\ref{a1}), (\ref{a2}) and
(\ref{a3}) and it equals to
\begin{equation}
z_{eff}=\frac{16}{3\pi}(a_1+a_2+a_3)/e^2.
\end{equation}
Note that, if we take $z_{eff}=1$ we obtain the point of maximum
of the integrand in the equation of the $pp$ reaction rate
constant.

To obtain the rate of reactions, the rate constant must be
multiplied by the density of the reactants,
\begin{equation}\label{rate-n}
{\cal{R}}_{pep}={\cal{K}}_{pep}n_p^2n_e,
\end{equation}
where $n_e$ and $n_p$ are the numbers of electrons and protons in
the unit of volume.

All solar parameters like temperature and densities of protons and
electrons vary with the radius of the interior of the Sun. The
results for the neutrino fluxes presented by Bahcall \textit{et
al.} \cite{bahcall-sab2} show that the fluxes from the $pp$ and
$pep$ reaction are not sensitive to the considered solar models.
Therefore, to calculate of the rates of the $pp$ and $pep$
reaction we apply the BS05(OP) model \cite{www-sns-ias-edu}.

\begin{figure}
\epsfig{file=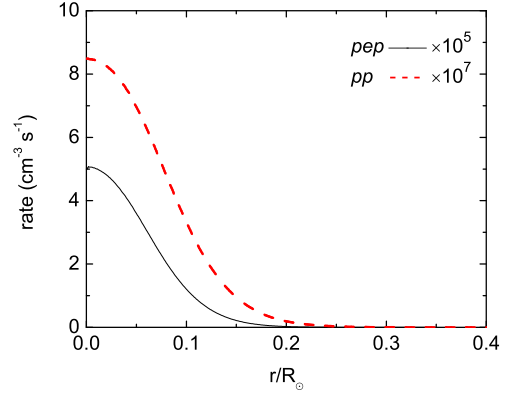,width=7.8cm} \caption{(Color online) The
$pep$ (solid line) and $pp$ (dashed line) rates versus  the radius
of the interior of the Sun. The results are presented for the
Gauss $NN$ potentials. The behavior of the rate curves for the
Yukawa $NN$ potentials is very close to the displayed curves. The
results are presented in units of $10^7\,(pp)\,\,{\rm{and}}
\,\,10^5\,(pep)\,\,\rm{cm}^{-3}\rm{s}^{-1}$. $R_{\odot}$ is the
radius of the Sun. \label{fig3}}
\end{figure}

Figure {\ref{fig3}} shows the rate of the $pp$ and $pep$ reaction
as a function or the solar interior radius for the Gauss $NN$
potentials. We obtain the same behavior in the case of the Yukawa
$NN$ potentials. We see that the rate of the $pep$ reactions is
more than a hundred times less than the rate of the $pp$ reaction
for the whole distance from the center of the Sun, and reactions
occur in the core of the Sun where the temperature and density are
higher.

Integrating the reaction rate (\ref{rate-n}) over the volume  of
the Sun we find the total flux of neutrinos emitted by the Sun.
Dividing this total flux by the area of the sphere of the radius
of one astronomical unit (AU) we obtain the neutrino flux $\Phi$
passing through a unit area of the Earth surface. The results of
calculations of the neutrino fluxes are presented in Table
\ref{tab1}.
\begin{table}[!t]
\caption{Predicted fluxes $\Phi_{pp}$ and $\Phi_{pep}$ (without
survival probability), in units of $10^{10}\,(pp), 10^8\,(pep)
\,\,\rm{cm^{-2}s^{-1}}$.}
\begin{ruledtabular}
\begin{tabular}{lcccl}
{\bf Standard Solar}&\,${\boldsymbol\Phi_{pp}}$\,
&\,${\boldsymbol\Phi_{pep}}$\,&${\boldsymbol\Phi_{pp}/\boldsymbol\Phi_{pep}}$&{\bf
   References}\\
   {\bf Model} & &&&\\
 BS05(OP)&6.20 &2.04&304 & our results \\
 &&&&with Gauss\\
 &&&& potential\\
 BS05(OP)&6.05 &1.99&304 & our results \\
 &&&&with Yukawa\\
 &&&& potential\\
 BP04(Yale)&5.94&1.40&424 & \,\,\,\, \cite{bahcall-sab1} \\
 BP04(Garching)&5.94&1.41&421 &\,\,\,\, \cite{bahcall-sab1} \\
 BS04&5.94&1.40&424 & \,\,\,\, \cite{bahcall-sab1} \\
 BS05(14N)&5.99&1.42&421 & \,\,\,\, \cite{bahcall-sab1} \\
 BS05(OP)&5.99&1.42&421 & \,\,\,\, \cite{bahcall-sab1} \\
 BS05(AGS,OP)&6.06&1.45&418 & \,\,\,\, \cite{bahcall-sab1} \\
 BS05(AGS,OPAL)&6.05&1.45&417 &\,\,\,\, \cite{bahcall-sab1} \\
 \end{tabular}
\end{ruledtabular}
\label{tab1}
\end{table}
Note that the results for the neutrino flux from the $pp$ reaction
are close to the results obtained by Bahcall \textit{et al.} while
there are differences between the fluxes from the $pep$ reactions.

The Borexino collaboration announced the results of the neutrino
flux measurement: $\Phi_{pep}=(1.6\pm 0.3)\times 10^8
{\rm{cm}}^{-2} {\rm{s}}^{-1}$ \cite{borexino}. Taking into account
that the survival probability of the neutrino (due to the neutrino
oscillation) in the $pep$ reaction predicted by the Borexino
collaboration equals  $P=0.62 \pm 0.17$ at 1.44 MeV
\cite{borexino}, we find that the neutrino flux at 1 AU should be
equal to $\boldsymbol{\Phi_{pep}= (1.27\pm ± 0.35)\times 10^8\,\,
{\rm{cm}}^{-2}{\rm{s}}^{-1}}$ for the Gauss potential, and
$\boldsymbol{\Phi_{pep}= (1.24\pm ± 0.34)\times 10^8\,\,
{\rm{cm}}^{-2}{\rm{s}}^{-1}}$ for the Yukawa potential.

We note that our results for the neutrino fluxes from the $pep$
reaction obtained by taking into account the survival probability
lie within the confidence interval of the experimental data. At
the same time, the Bahcall results are out of this limit at all
fluxes listed  in Table \ref{tab1} if they are multiplied by the
same survival probability of the neutrino. Comparing our
calculated low-energy parameters for the Gauss and Yukawa
potentials, we see that they differ by 2\% to7\%, and the neutrino
fluxes  from the $pp$ have a 2\% difference,  too, and the results
of $\Phi_{pp}$ of Bahchall \textit{et al.} shown in Table
\ref{tab1} for all SSMs are not more than 2\% to 4\% from our
calculated $\Phi_{pp}$. Therefore, we may conclude that the
dependence of the results on the type of $NN$ potentials is very
weak. However, the difference in the results obtained by ourselves
and Bahcall \textit{et al.} can reach up to 39\% to 45\%, which
means strong sensitivity to the choice of the wave functions of
the initial three-body state of the $pep$ system.

\section{Conclusion}\label{fin}
In the framework of the three-body approach the probability of the
process $pep\to d+\nu_e$ under  conditions of the solar core has
been found. The rate of the above process and neutrino flux are
found and compared with the Borexino experiment and previous
calculations. The value of the neutrino flux obtained from the
$pep$ reaction in the three-body treatment appeared to be $\sim
40$\% larger as compared to the Bahcall \textit{et al.} value.
This can be understood as a correct description of the movement of
the electron producing the screening effect between protons. In
this work we have introduced the astrophysical $S$ factor for the
three-body reaction which is an analog of the $S$ factor
introduced for binary processes. To discriminate between different
star models on the basis of our results, it is necessary to
essentially reduce experimental errors in the above experiment.

\section*{Acknowledgments} The authors are
very grateful to Prof. O.\,A.~Zaimidoroga for fruitful discussions
and useful comments.

\appendix
\section{Schr\"odinger equation}\label{app1}
The Hamiltonian of the $pep$ system is
\begin{equation}\label{hamiltonian}
H=H_0+V^N_{23}+V^C_{123},
\end{equation}
where $H_0$ is kinetic energy operator:
\begin{equation}\label{kin}
H_0=-\frac{\hbar^2}{2\mu_{23}}\Delta_{23}-
\frac{\hbar^2}{2M_{1(23)}}\Delta_{1(23)},
\end{equation}
subscript $1$ means electron, while $2$ and $3$ correspond to two
protons; $\mu_{23}$  is the reduced mass of two protons, while
$M_{1(23)}$ is the reduced mass of the system of two protons and
electron;
\begin{equation}\label{vcoul}
V^C_{123}=V^C_{12}+V^C_{23}+V^C_{31}
\end{equation}
is the sum of the Coulomb potentials in the $pep$ system, and
$V^N_{23}$ is the nuclear potential of interaction between two
nucleons.

The wave function of the initial state of the $pep$ system  is the
eigenfunction of the Hamiltonian (\ref{hamiltonian}), which has a
continuous spectrum of energy and satisfies the Schr\"odinger
equation
\begin{equation}\label{sch eq}
H\Psi({\boldsymbol{r_1},\boldsymbol{r_2},\boldsymbol{r_3}})=E\Psi({\boldsymbol{r_1},\boldsymbol{r_2},\boldsymbol{r_3}}).
\end{equation}

Let us define the Jacobi coordinates:
\begin{eqnarray} \label{jacob}
{\boldsymbol{x_i}}&=&\sqrt{\frac{m_j
m_k}{(m_j+m_k)\mu_{23}}}({\boldsymbol{r_j}}-{\boldsymbol{r_k}})\nonumber\\
{\boldsymbol{y_i}}&=&\sqrt{\frac{m_i(m_j+m_k)}{(m_1+m_2+m_3)\mu_{23}}}\times\nonumber\\\
&&\Bigl(-{\boldsymbol{r_i}}+\frac{m_j{\boldsymbol{r_j}}+m_k{\boldsymbol{r_k}}}
{m_j+m_k}\Bigr),
\end{eqnarray}
where $\mu_{23}=m_2m_3/(m_2+m_3)$ is the reduced mass of two
protons, indices \textit{i\,j\,k}=123, 231, or 312 and ${m_i}$
($\boldsymbol{r_i}$)  is the mass (coordinate) of particle $i$. In
introducing coordinates the Hamiltonian $H_0$ of the free motion
of particles can be written as
\begin{eqnarray}\label{H0}
H_0&=&-\frac{\hbar^2}{\mu_{23}}\Bigl(\Delta_{{\boldsymbol{x_1}}}+\Delta_{{\boldsymbol{y_1}}}\Bigr)\equiv
-\frac{\hbar^2}{\mu_{23}}\Bigl(\Delta_{{\boldsymbol{x_2}}}+\Delta_{{\boldsymbol{y_2}}}\Bigr)\nonumber\\
&\equiv&-\frac{\hbar^2}{\mu_{23}}\Bigl(\Delta_{{\boldsymbol{x_3}}}+\Delta_{{\boldsymbol{y_3}}}\Bigr).
\end{eqnarray}
We define the square of the hyperradius and  hyperangle as:
\begin{eqnarray}
\rho^2&=&x_1^2+y_1^2\equiv x_2^2+y_2^2\equiv
x_3^2+y_3^2,\,\,\,\,\,\,\, \rho\in [0,\infty),\nonumber\\
x_i&=&\rho\cos\alpha_i,\,\,\,y_i=\rho\sin\alpha_i, \,\,\,\,\,
\alpha_i\in [0,\pi/2].\nonumber
\end{eqnarray}
In the variables of hyperradius and a set of angles
$\Omega_i=({\boldsymbol{\hat{x}_i,\hat{y}_i}},\alpha_i)$
(${\boldsymbol{\hat{x}_i}}$ and ${\boldsymbol{\hat{y}_i}}$ are
unit vectors determining the azimuthal and polar angles)), we can
rewrite the operator $H_0$ as
\begin{eqnarray}\label{H02}
H_0=-\frac{\hbar^2}{2\mu_{23}}\Bigl(\frac{\partial^2}{\partial\rho^2}+
\frac{5}{\rho}\frac{\partial}{\partial\rho}-\frac{1}{\rho^2}K^2\bigl(\Omega_i\bigr)\Bigr),\\
\nonumber
\end{eqnarray}
where the operator $K^2\bigl(\Omega_i\bigr)$ is
\begin{eqnarray}\label{K2}
K^2\bigl(\Omega_i\bigr)&=&-\frac{\partial^2}{\partial\alpha_i^2}-4\cot
2\alpha_i\frac{\partial}{\partial\alpha_i}+\nonumber\\
&&\frac{1}{\cos^2\alpha_i}l^2({\boldsymbol{\hat{x}_i})}
+\frac{1}{\sin^2\alpha_i}l^2({\boldsymbol{\hat{y}_i})},
\end{eqnarray}
where $l^2=-\Delta_{\theta,\varphi}$ is an angular part of the
Laplace operator.

The hyperspherical function is defined as the solution of the
equation \cite{djibuti}
\begin{equation}\label{hypersph}
K^2\Phi(\Omega_i)=K(K+4)\Phi(\Omega_i),\,\,\,\,\, K=0,1,2,3,\dots
\end{equation}
The quantum number $K$ is called hypermoment, the eigenfunction
$\Phi(\Omega_i)$ is
\begin{align}\label{eigen}
&\Phi(\Omega_i)\equiv\Phi^{l_{x_i}l_{y_i}}_{KLM}(\Omega_i)\nonumber\\
&=\sum\limits_{m_{x_i}m_{y_i}}\bigl(l_{x_i}m_{x_i}l_{y_i}m_{y_i}\mid
LM\bigr)\Phi^{l_{x_i}l_{y_i}m_{x_i}m_{y_i}}_{KLM}(\Omega_i),
\end{align}
where the function
$\Phi^{l_{x_i}l_{y_i}m_{x_i}m_{y_i}}_{KLM}(\Omega_i)$ is
\begin{eqnarray}\label{eigen1}
\Phi^{l_{x_i}l_{y_i}m_{x_i}m_{y_i}}_{KLM}(\Omega_i)=N_K^{l_{x_i}l_{y_i}}(\cos\alpha_i)^{l_{x_i}}(\sin\alpha_i)^{l_{y_i}}\times\nonumber\\
P_n^{l_{y_i}+1/2,l_{x_i}+1/2}(\cos2\alpha_i)Y_{l_{x_i}m_{x_i}}({\boldsymbol{\hat{x}_i}})Y_{l_{y_i}m_{y_i}}({\boldsymbol{\hat{y}_i}});\,\,\,\,\,\,\,
\end{eqnarray}
$\bigl(l_{x_i}m_{x_i}l_{y_i}m_{y_i}\mid LM\bigr)$ is the
Clebsch-Gordon coefficient; $Y_{l_x m_x}({\boldsymbol{\hat{x}}})$
is the spherical function;
\begin{equation}\label{NKn}
N_K^{l_{x_i}l_{y_i}}=\sqrt{\frac{2n!(K+2)(n+l_{x_i}+l_{y_i}+1)!}{\Gamma\bigl(n+l_{x_i}+3/2\bigr)
\Gamma\bigl(n+l_{y_i}+3/2\bigr)}},
\end{equation}
$n=(1/2)\bigl(K-l_{x_i}-l_{y_i}\bigr)$ must be an integer number;
$P_n^{l_xl_y}$ is Jacobi polynomial.

Let $\bf{p}$ and $\bf{q}$ be conjugate momenta to the coordinates
$\boldsymbol{x}$ and $\boldsymbol{y}$. Then determining square of
the wave number $\kappa^2=2\mu_{23}E/\hbar^2$ ($E$ is the total
energy of the $pep$ system in the c.m. frame) for the $pep$ system
in the continuous state, removing a free motion of the center of
mass of the $pep$ system from Eq. (\ref{sch eq}) and using the
following expansion of the $pep$ wave function of a continuous
spectrum over the hyperharmonics functions
\begin{eqnarray}\label{expansion}
\Psi_{\bf{p},\bf{q}}({\bf{x},\bf{y}})=(2\pi)^3\sum\limits_{Kl_{x}l_{y}LM}i^K
\frac{U_{KL}^{l_{x}l_{y}}(\rho)}{(\kappa\rho)^2}\times
\nonumber\\
\Phi^{l_{x}l_{y}}_{KLM}(\Omega_\rho)\Phi^{*l_{x}l_{y}}_{KLM}(\Omega_\kappa)
\end{eqnarray}
we get the radial Schr\"odinger equation:
\begin{eqnarray}\label{rad eq}
\frac{d^2U_{KL}^{l_xl_y}(\rho)}{d\rho^2}+\frac{1}{\rho}\frac{dU_{KL}^{l_xl_y}(\rho)}{d\rho}-\Bigl[\frac{(K+2)^2}{\rho^2}-
\kappa^2\Bigr]U_{KL}^{l_xl_y}(\rho)\nonumber\\
=\sum\limits_{K' l'_x l'_y}{\cal{V}}_{KK' LM}^{l_xl_y;l'_x
l'_y}(\rho)U_{K'L}^{l'_x
l'_y}(\rho),\,\,\,\,\,\,\,\,\,\,\,\,\,\,\,\,\,\,\,\,\,\,\,\,\,\,\,\,\,\,
\end{eqnarray}
where superscript $(^*)$ denotes taking complex conjugate of the
function, and $\Phi^{l_{x}l_{y}}_{KLM}(\Omega_\rho)\equiv
\Phi^{l_{x}l_{y}}_{KLM}(\Omega_1)$, where
$\Phi^{l_{x}l_{y}}_{KLM}(\Omega_\kappa)$ is defined on a
hypersphere of unit radius in a six dimensional momentum space.
The matrix element ${\cal{V}}_{KK' LM}^{l_xl_y;l'_x l'_y}(\rho)$
equals to
\begin{align}\label{matrixV}
{\cal{V}}_{KK' LM}^{l_xl_y;l'_x
l'_y}(\rho)=\frac{2\mu_{23}}{\hbar^2}\int\Phi^{*l_{x}l_{y}}_{KLM}(\Omega_\rho)\times\,\,\,\,\,\,\,\,\,\,\,\,\,\,\,\nonumber\\
\Bigl(V_{23}^N(\mid{\boldsymbol{x_1}\mid})+V_{23}^C(\mid{\boldsymbol{x_1}\mid})+V_{31}^C(\mid{\boldsymbol{x_2}\mid})+\\
V_{12}^C(\mid{\boldsymbol{x_3}\mid})\Bigr)\Phi^{l'_{x}l'_{y}}_{K'LM}(\Omega_\rho)d\Omega_\rho\,\,\,\,\,\,\,\,\,\,\,\,\,\,\nonumber
\end{align}
The system of Eq. (\ref{rad eq}) is a system of the linked one
dimensional equations which must satisfy the boundary conditions
depending on the particular physical situation.

It is easy to show that, near $\rho=0$, regular solutions of
(\ref{rad eq}) must behave as $\rho^{K+2}$.  If we omit the
nondiagonal terms in the equations, solutions at $\rho\to\infty$
have the asymptotic behavior as a superposition of the regular and
irregular Coulomb functions \cite{djibuti}.

\end{document}